\documentclass[9pt,conference,a4paper]{IEEEtran}
\usepackage[latin1]{inputenc}
\usepackage{amssymb}
\usepackage{latexsym}
\usepackage{mathrsfs}
\usepackage{amsfonts}
\usepackage{amsbsy}
\usepackage{amsmath}
\usepackage{times}
\usepackage{epsfig,epsf,times,amssymb,amsmath}
\title{Low-Complexity Constrained Constant Modulus SG-based Beamforming Algorithms with Variable Step Size}
\author{%
\fontsize{10}{10}\selectfont\itshape {Lei Wang, Yunlong Cai, and Rodrigo C. de Lamare}\\
\normalfont Communications Research Group, Department of Electronics, The University of York, York, YO10 5DD, UK\\
Email: lw517@york.ac.uk,   yc521@york.ac.uk,   rcdl500@ohm.york.ac.uk}

\begin{document}
\maketitle
\begin{abstract}
In this paper, two low-complexity adaptive step size algorithms are investigated for blind adaptive beamforming. Both of them
are used in a stochastic gradient (SG) algorithm, which employs the constrained constant modulus (CCM) criterion as the design
approach. A brief analysis is given for illustrating their properties. Simulations are performed to compare the performances of
the novel algorithms with other well-known methods. Results indicate that the proposed algorithms achieve superior performance,
better convergence behavior and lower computational complexity in both stationary and non-stationary environments.\\


\end{abstract}

\section{\bfseries Introduction}
In recent years, many adaptive filtering algorithms have been used for beamforming, which is a promising and widely investigated
technology for rejecting interference and improving the performance of high capacity mobile communications systems
\cite{Falletti}. Many methods have been presented in different communication systems \cite{Gershman}-\cite{Jian}. In contrast to
fixed beamforming techniques, an adaptive beamformer has the ability of rejecting interference and pointing its mainbeam in the
desired direction with the change of scenarios. Blind adaptive beamforming, which is intended to form the array direction
response without knowing users' information beforehand, is an important topic that deals with interference cancellation,
tracking improvement and complexity reduction.

The blind adaptive SG method, which is commonly employed in the blind adaptive beamforming area, is a well-known technique for
solving optimization problems with different criteria, e.g., minimum mean squared error (MMSE) \cite{Madhow}, minimum variance
(MV) \cite{Honig} and constant modulus (CM) \cite{Johnson}. The MV algorithm is a computational efficient approach to estimate
the input covariance matrix. The results in \cite{Honig} prove that the MV criterion leads to a solution identical to that
obtained from the minimization of the mean squared error (MSE). The CM algorithm for beamforming exploits the low modulus
fluctuation exhibited by communications signals using constant modulus constellations to extract them from the array input. It
is well known that the performance of the CM method is superior to that of the MV. A disadvantage of both two methods is that
they are sensitive to the step size. The small value of the step size will lead to slow convergence rate, whereas a large one
will lead to high misadjustment or even instability. Besides, the CM cost function may have local minima, and CM receivers do
not have closed-form solutions. Xu and Liu \cite{Xu} developed a SG algorithm on the basis of the CCM technique to sort out the
local minimum problem and obtain the global minima. But the problem incurred by the step size cannot be solved properly.

For accelerating the convergence, recursive least squares (RLS) algorithms were introduced by Xu and Tsatsanis using the
constrained minimum variance (CMV) criterion \cite{Tsatsanis} and then developed by de Lamare and Sampaio-Neto with the CCM
approach \cite{de Lamare}. The latter, which improves the performance significantly, optimizes a quadratic cost function based
on the CM criterion subject to linear constraints for the array weight adaptation. Combining with the constrained condition,
this method reaches an optimal solution. Nevertheless, the RLS based beamformers cannot avoid complicated computations caused by
the required correlation matrix inversion.

Comparing SG algorithms, which represent simple and low-complexity solutions but subject to slow convergence, with RLS methods,
which possess fast convergence but high computational load, it is preferable to adopt SG beamformers due to complexity and cost
issues. For this reason, the improvement of blind SG techniques is an important topic and has been investigated for several
decades. In this research area, the works in \cite{Honig} and \cite{Tsatsanis} employ standard SG methods with fixed step size
(FSS) that are not efficient with respect to convergence and steady-state performance. The performance of the beamformer is
strongly dependent on the choice of the step size \cite{Haykin}. It reflects a tradeoff between misadjustment and the
convergence rate. Actually, the communication systems are non-stationary environments, which make it very difficult to
predetermine the step size. It is necessary to make the step size track the change of the system automatically and so obtain
good convergence behavior. Previous researches have focused on this aim and some good results have been reported. The adaptive
step size (ASS) mechanism was employed in both the MV \cite{Mathews}, \cite{Kushner} and the CM \cite{Yuvapoositanon} criteria
for improving the performance. Because of requiring an additional update equation for the gradient of the weight vector with
respect to the step size, which increases the extra computational load, the applications of these algorithms are limited. The
authors of \cite{Rodrigo} propose two novel variable step size mechanisms for MV algorithms. The simulation experiments show
that the new mechanisms are superior to previously reported methods and have a reduced complexity.

This paper proposes two blind CCM beamformers based on two novel adaptive step size mechanisms. The origins of these mechanisms
can be traced back to the works of \cite{Kwong} and \cite{Aboulnasr} where low-complexity adaptive step size mechanisms were
developed for LMS algorithms. In contrast to \cite{Rodrigo}-\cite{Aboulnasr}, the mechanisms here are designed for CCM
algorithms, since it is well-known that they outperform the CMV algorithms \cite{Xu}. Because of this reason, the simulations
here just compare the algorithms related to the CCM criterion. The additional number of operations of the proposed algorithms
does not depend on the number of sensor elements. In addition, the results are presented for stationary and non-stationary
environments, proving that the new mechanisms could reach better performance and faster convergence behavior than those of
previous methods.

The remaining of this paper is organized as follows. In the next section, we present a system model for smart antennas. Based on
this model, the blind adaptive CCM beamformer design using the SG method is presented in Section III. In Section IV, the
proposed adaptive step size mechanisms are derived. Simulation results are provided in Section V, and conclusions are drawn in
Section VI.

\section{\bfseries SYSTEM MODEL}

In order to describe the system structure, let us make two simplifying assumptions for the transmitter and receiver models
\cite{Dimitris}. First, the propagating signals are assumed to be produced by point sources; that is, the size of the source is
small with respect to the distance between the source and the sensors that measure the signal. Second, the sources are assumed
to be in the "far field," namely, at a large distance from the sensor array, so that the spherically propagating wave can be
reasonably approximated with a plane wave. Besides, we assume a lossless, nondispersive propagation medium, i.e., a medium that
does not attenuate the propagating signal further and the propagation speed is uniform so that the waves travel smoothly.

\begin{figure}[ht]
\begin{center}
\includegraphics[angle=0,
width=0.45\textwidth,height=0.28\textheight]{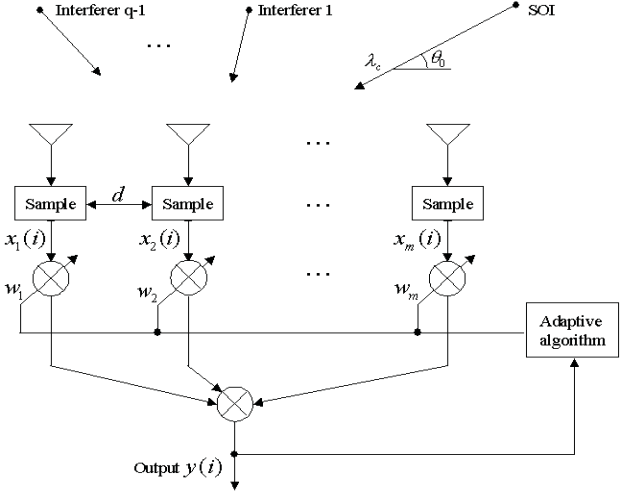}
\end{center} \caption{\label{fig:model4} Adaptive beamforming structure for ULA.}
\end{figure}

Let us consider the adaptive beamforming scheme in Fig. \ref{fig:model4} and suppose that $q$ narrowband signals impinge on the
uniform linear array (ULA) of $m$ ($q \leq m$) sensor elements from the sources with unknown directions of arrival (DOAs)
$\theta_{0}$,\ldots,$\theta_{q-1}$. The $i$th snapshot's vector of sensor array outputs can be modeled as \cite{Stoica}

\begin{equation} \label{1}
\centering {\boldsymbol x}(i)={\boldsymbol A}({\boldsymbol {\theta}}){\boldsymbol s}(i)+{\boldsymbol n}(i),~~~ i=1,\ldots,N
\end{equation}
where $\boldsymbol{\theta}=[\theta_{0},\ldots,\theta_{q-1}]^{T}\in\mathcal{C}^{q \times 1}$ is the vector of the unknown signal
DOAs, ${\boldsymbol A}({\boldsymbol {\theta}})=[{\boldsymbol a}(\theta_{0}),\ldots,{\boldsymbol
a}(\theta_{q-1})]\in\mathcal{C}^{m \times q}$ is the complex matrix composed of the signal direction vectors ${\boldsymbol
a}(\theta_{k})=[1,e^{-2\pi j\frac{d}{\lambda_{c}}cos{\theta_{k}}},\ldots,e^{-2\pi
j(m-1)\frac{d}{\lambda_{c}}cos{\theta_{k}}}]^{T}\in\mathcal{C}^{m \times 1},~~~(k=0,\ldots,q-1)$, where $\lambda_{c}$ is the
wavelength and $d=\lambda_{c}/2$ is the inter-element distance of the ULA, ${\boldsymbol s}(i)\in \mathcal{R}^{q\times 1}$ is
the real value vector of the source data, ${\boldsymbol n}(i)\in\mathcal{C}^{m\times 1}$ is the complex vector of white sensor
noise, which is assumed to be a zero-mean spatially and white Gaussian process, $N$ is the number of snapshots, and
$(\cdot)^{T}$\ stands for the transpose. The output of a narrowband beamformer is given by

\begin{equation} \label{2}
\centering y(i)={\boldsymbol w}(i)^H {\boldsymbol x}(i)
\end{equation}
where ${\boldsymbol w}(i)=[w_{1}(i),\ldots,w_{m}(i)]^{T}\in\mathcal{C}^{m\times 1}$ is the complex weight vector, and
$(\cdot)^{H}$ stands for the Hermitian transpose.

\section{\bfseries Blind Adaptive CCM Algorithms}

The purpose of SG algorithms is to get an acceptable output performance and reduce the complexity load by avoiding the
correlation matrix estimation and inversion. We describe the CCM algorithm on the basis of the SG method, which is called
CCM-SG.

For the CCM-SG algorithm, we consider the cost function as the expected deviation of the squared modulus of the array output to
a constant, say 1. The CCM cost function is simply a positive measure of the average amount that the beamformer output deviates
from the unit modulus condition \cite{Jian}. By using the constraint condition, the cost function of CCM-SG can be expressed as

\begin{equation}\label{3}
\begin{split}
\centering
&J_{CM}=(|y(i)|^2-1)^2,~~i=1,\ldots,N\\
&\textrm{subject~to}~~~{\boldsymbol w}^{H}(i){\boldsymbol a}(\theta_{0})=1
\end{split}
\end{equation}
where $\boldsymbol a(\theta_{0})$ denotes the steering vector of the desired signal. The constrained optimization means that the
technique minimizes the contribution of the undesired interferences while maintaining the gain along the look direction to be
constant.

The SG blind adaptive algorithm optimizes the Lagrangian cost function described by

\begin{equation}\label{4}
L_{CCM}=(|y(i)|^2-1)^2+\lambda({\boldsymbol w}^{H}(i){\boldsymbol a}(\theta_{0})-1)
\end{equation}
where $\lambda$ is a scalar Lagrange multiplier. The solution can be obtained by setting the gradient terms of (\ref{4}) with
respect to $\boldsymbol w$(i) equal to zero and using the constraint. Thus, we obtain

\begin{equation}\label{5}
\begin{split}
{\boldsymbol w}(i+1)&={\boldsymbol w}(i)-\mu(|y(i)|^2-1) y^{\ast}(i)\\&\quad \cdot[{\boldsymbol x}(i)-{\boldsymbol
a}^{H}(\theta_{0}){\boldsymbol x}(i){\boldsymbol a}(\theta_{0})]
\end{split}
\end{equation}
where $\mu$ here is the step size, which is a fixed value for FSS and a variable value for ASS and $\ast$ denotes complex
conjugate.

Because of the shortcomings introduced before for both FSS and ASS algorithms, it is necessary to develop other methods for
improving the performance of SG method.

\section{\bfseries Proposed Adaptive Step Size Mechanisms}

In this section, two novel adaptive step size methods are described for adjusting the step size following the change of the
communication system. The step size adjustment is controlled by the square prediction error, which means that a large error will
cause the step size to increase for providing fast tracking while a small error will result in a decrease of step size to yield
smaller misadjustment. The computational complexity is not a problem in these mechanisms.

\subsection{\bfseries Modified Adaptive Step Size (MASS) Mechanism}

The first proposed algorithm based on the MASS mechanism employs the prediction error and uses the update rule

\begin{equation}\label{6}
\mu(i+1)=\alpha \mu(i)+\gamma(|y(i)|^{2}-1)^{2}
\end{equation}
where $0<\alpha<1$, $\gamma>0$ and $y(i)$ is the same as that in (\ref{2}). The rationale for the MASS is that for large
prediction error the algorithm will make the step size increase to track the change of the system whereas a small error will
result in a decrease of the step size. The parameter $\gamma$ is an independent variable for controlling the prediction error
and scaling it at different levels. It is worth pointing out that the step size $\mu(i+1)$ should be limited in a range as
follows

\begin{equation}\label{7}
\mu(i+1)=\left\{ \begin{array}{ccc}
                  \mu_{max} & \textrm {if}~\mu(i+1)>\mu_{max}\\
                  \mu_{min} & \textrm {if}~\mu(i+1)<\mu_{min}\\
                  \mu(i+1) & \textrm {otherwise}\\
                  \end{array}\right .
\end{equation}
where $0<\mu_{min}<\mu_{max}$. The constant $\mu_{min}$ is chosen as a compromise between the satisfying level of steady-state
misadjustment and the required minimum level of tracking ability while $\mu_{max}$ is normally selected close to the point of
instability of the algorithm for providing the maximum convergence speed. The MASS is the result of several attempts to devise a
simple and effective mechanism.

\subsection{\bfseries Time Averaging Adaptive Step Size (TAASS) Mechanism}

The second mechanism, which is called TAASS, uses a time average estimate of the correlation of $(|y(i)|^{2}-1)$ and
$(|y(i-1)|^{2}-1)$. The update rule is

\begin{equation}\label{8}
\mu(i+1)=\alpha \mu(i)+\gamma v^{2}(i)
\end{equation}
where $v(i)=\beta v(i-1)+(1-\beta)(|y(i)|^{2}-1)(|y(i-1)|^{2}-1)$ and $0<\beta<1$. The limits on $\mu(i+1)$, $\alpha$ and
$\gamma$ are similar to those of the MASS algorithm. The exponential weighting parameter $\beta$ governs the averaging time
constant, namely, the quality of the estimation. Previous samples, in stationary environments, contain information that is
related to determine an accurate measure for the proximity of the adaptive beamformer coefficients to the optimal ones. Here,
$\beta$ should be close to $1$. For non-stationary environments, the time averaging window should be small for deleting the deep
past data and leaving space for the current statistics adaption, so, $\beta<1$.

There are two objectives for using $v(i)$ here. First, it rejects the effect of the uncorrelated noise sequence on the step-size
update \cite{Aboulnasr}. In the beginning, because of scarcity of transmitters' information, the error correlation estimate
$v^{2}(i)$ is large and so $\mu(i)$ is large to increase the convergence rate and to track the change of input data. As it
approaches the optimum, $v^{2}(i)$ is very small , resulting in a small step size for ensuring low misadjustment near optimum.
Second, the error correlation is generally a good measure of the proximity to the optimum.

\subsection{\bfseries Computational Complexity and Convergence Analysis}

In this part, both the computational complexity and the convergence behavior of the proposed mechanisms based on the CCM
criterion are investigated.

\subsubsection{Computational Complexity}
The computational complexities of the proposed MASS and TAASS mechanisms are analyzed. It is well known that the computational
complexity of the ASS algorithm is a linear monotonic increasing function of the number of sensor elements (in AWGN model),
i.e., the complexity will increase following the increase of the number of sensor elements. Therefore, the computational
complexity becomes very large if the array size is big.

An important feature of the proposed algorithms is that they only require a few fixed number of operations for updating the step
size compared with that of the ASS method, which is proportional to the number of sensor elements. The additional computational
complexity of the proposed adaptive step size mechanisms is listed in Table \ref{tab:complexity}.

\begin{table}[!t]
\centering

    \caption{Additional Computational Complexity of Proposed Algorithms}     
    \label{tab:complexity}

    \begin{small}
    \begin{tabular}{|c|l|c|l|c|l|}
    \hline
    {Proposed} & \multicolumn{2} {c|} {Number of operations per snapshot} \\
    \cline{2-3}
    {Algorithms} &\quad {Additions}\quad\quad  &{Multiplications}  \\
    \hline
    \bfseries MASS      &\quad\quad\quad  1   & 3  \\
    \hline
    \bfseries TAASS     &\quad\quad\quad   2   & 6  \\
    \hline
    \end{tabular}
    \end{small}
\end{table}

\subsubsection{Convergence Analysis}

Considering the space limitation, we just give the range of the step size for convergence. For further analysis, we assume that
$\mu(i)$ varies slowly around its mean value. This assumption is approximately true if $\gamma$ is small and $\alpha$ closes to
one, which will be shown in the simulations. Under this condition, according to \cite{Rodrigo}, we have

\begin{equation}\label{9}
\begin{split}
&E[\mu(i)(|y(i)|^{2}-1)y^{\ast}(i)[\boldsymbol x(i)-\boldsymbol a^{H}(\theta_{0})\boldsymbol x(i)\boldsymbol a(\theta_{0})]]\\
&=E[\mu(i)]E[(|y(i)|^{2}-1)y^{\ast}(i)[\boldsymbol x(i)-\boldsymbol a^{H}(\theta_{0})\boldsymbol x(i)\boldsymbol a(\theta_{0})]]
\end{split}
\end{equation}

and

\begin{equation}\label{10}
E[\mu(i)(|y(i)|^{2}-1)\boldsymbol x(i)\boldsymbol x^{H}(i)]\boldsymbol w(i)=E[\mu(i)]\boldsymbol R_{CCM}\boldsymbol w(i)
\end{equation}
where $\boldsymbol R_{CCM}=E[(|y(i)|^{2}-1)\boldsymbol x(i)\boldsymbol x^{H}(i)]\in\mathcal C^{m\times m}$.

Now, the weight vector update equation (\ref{5}) of the blind adaptive CCM beamformer can be written as

\begin{equation}\label{11}
\boldsymbol w(i+1)=(\boldsymbol I-\mu(i)(|y(i)|^{2}-1)\boldsymbol v(i)\boldsymbol x^{H}(i))\boldsymbol w(i)
\end{equation}
where $\boldsymbol v(i)=(\boldsymbol I-\boldsymbol a(\theta_{0})\boldsymbol a^{H}(\theta_{0}))\boldsymbol x(i)\in\mathcal
C^{m\times 1}$.

Define the weight error vector $\boldsymbol e_{w}(i)$ and substitute (\ref{11}) into the expression, we get

\begin{equation}\label{12}
\begin{split}
\boldsymbol e_{w}(i+1)&=\boldsymbol w(i+1)-\boldsymbol w_{opt}\\
&=(\boldsymbol I-\mu(i)(|y(i)|^{2}-1)\boldsymbol v(i)\boldsymbol x^{H}(i))\boldsymbol e_{w}(i)\\
&~~~~-\mu(i)(|y(i)|^{2}-1)\boldsymbol v(i)\boldsymbol x^{H}(i)\boldsymbol w_{opt}
\end{split}
\end{equation}

Employing (\ref{9}) and (\ref{10}) and taking expectations on both sides of (\ref{12}), we get

\begin{equation}\label{13}
E[\boldsymbol e_{w}(i+1)]=(\boldsymbol I-E[\mu(i)]\boldsymbol R_{vx}(i))E[\boldsymbol e_{w}(i)]
\end{equation}
where $\boldsymbol R_{vx}(i)=E[(|y(i)|^{2}-1)\boldsymbol v(i)\boldsymbol x^{H}(i)]=(\boldsymbol I-\boldsymbol
a(\theta_{0})\boldsymbol a^{H}(\theta_{0}))\boldsymbol R_{CCM}$ and $\boldsymbol R_{vx}\boldsymbol w_{opt}=\boldsymbol 0$.
Therefore, $E[\boldsymbol w(i)]\rightarrow \boldsymbol w_{opt}$ or equivalently, $\lim_{i\rightarrow \infty}E[\boldsymbol
e_{w}(i)=\boldsymbol 0]$ represents the stable condition if and only if $\prod_{i=0}^{\infty}(\boldsymbol I-E[\mu(i)]\boldsymbol
R_{vx})\rightarrow 0$. Following the idea of the eigenstructure \cite{Haykin} with respect to the correlation matrix
$\boldsymbol R_{vx}$, the sufficient condition for (\ref{13}) to hold implies that

\begin{equation}\label{14}
0\leq E[\mu(\infty)]\leq \frac{2}{\lambda^{vx}_{max}}
\end{equation}
where $\lambda^{vx}_{max}$ is the maximum eigenvalue of $\boldsymbol R_{vx}$.

\section{\bfseries Simulations}

The performances of the proposed MASS and TAASS algorithms are compared with other existing algorithms, namely FSS and ASS, in
terms of output signal-to-interference-plus-noise ratio (SINR). An ULA containing $m=16$ sensor elements with half-wavelength
spacing is considered. The noise is spatially and temporally white Gaussian noise with power $\sigma_{n}^{2}=0.01$. For each
scenario, $K=1000$ iterations are used to get each simulated curve. In all simulations, the desired signal power is
$\sigma_{0}^{2}=1$. The BPSK modulation scheme is employed to modulate the signals.

\begin{figure}[!htb]
\begin{center}
\def\epsfsize#1#2{1.0\columnwidth}
\epsfbox{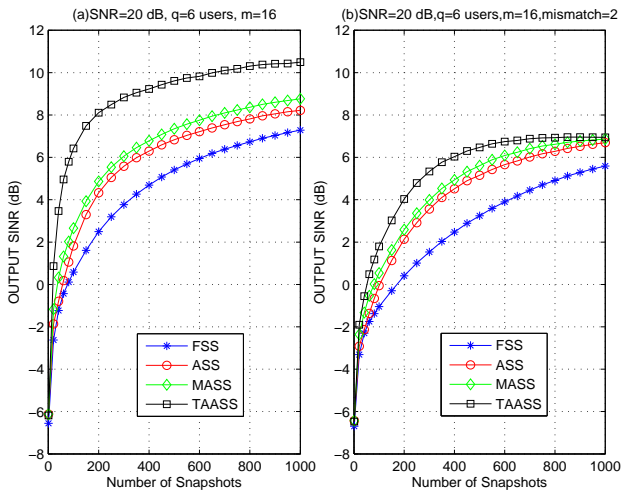} \caption{Output SINR versus number of snapshots
for (a) ideal steering vector condition (b) steering vector with
mismatch.} \label{fig:vssmismatch}
\end{center}
\end{figure}

Fig. \ref{fig:vssmismatch} includes two experiments. Fig.
\ref{fig:vssmismatch}(a) shows the output SINR of each method versus
the number of snapshots, whose total is $1000$ samples. There are
five interferers in the system, one interferer with $4$ dB above the
desired user's power level, one with the same power level of the
desired one and three with power $0.5$ dB lower than that of the
desired user. In this environment, the actual spatial signature of
the signal is known exactly. We set the first element of the initial
weight vector $\boldsymbol w(0)$ equals to the corresponding element
of steering vector of SOI $\boldsymbol a(\theta_{0})$. Other
parameters are optimized with $\alpha=0.98$, $\gamma=10^{-3}$,
$\mu_{0}=10^{-5}$, $\mu_{max}=10^{-4}$ and $\mu_{min}=10^{-6}$ for
MASS and $\alpha=0.98$, $\beta=0.99$,$\gamma=10^{-3}$,
$\mu_{0}=10^{-4}$, $\mu_{max}=3\times 10^{-4}$ and
$\mu_{min}=10^{-6}$ for TAASS. We claim that the parameters for the
FSS and ASS are tuned in order to minimize the performance, allowing
for a fair comparison with the proposed algorithms. The results show
that the proposed algorithms converge faster and have better
performances than the existing algorithms. The steering vector
mismatch scenario is shown in Fig. \ref{fig:vssmismatch}(b). We
assume that this steering vector error problem is caused by look
direction mismatch. The assumed DOA of the SOI is a constant value
$2^{o}$ away from the actual direction. Compared with Fig.
\ref{fig:vssmismatch}(a), Fig. \ref{fig:vssmismatch}(b) indicates
that the mismatch problem leads to a worse performance for all the
solutions. The convergence rate of all the methods reduces whereas
the devised algorithms are more robust to this mismatch, especially
for the TAASS approach, which reaches the steady-state very quickly.

\begin{figure}[!htb]
\begin{center}
\def\epsfsize#1#2{1.0\columnwidth}
\epsfbox{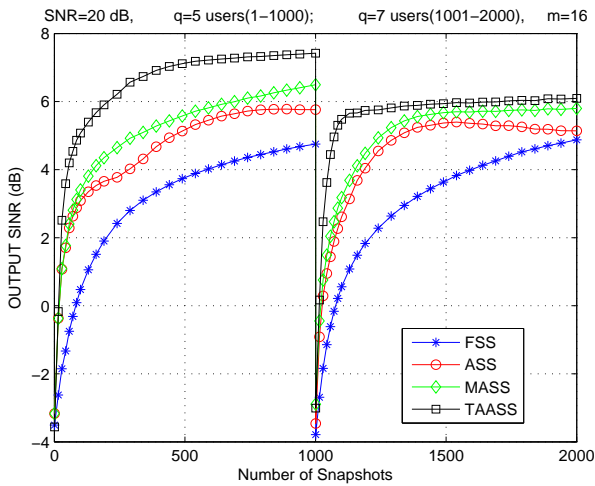} \caption{Output SINR in a scenario where
additional interferers suddenly enter and/or leave the system.}
\label{fig:vssmoreusers}
\end{center}
\end{figure}

In Fig. \ref{fig:vssmoreusers}, The system starts with $4$ interferers, two of which have the same power as that of the desired
signal and the rest of them with the power $0.5$ dB lower than the desired one. Two more users with one of them $2$ dB above the
desired user's power level and the other $0.5$ dB lower than that of the desired user, enter the system at $1000$ symbols. In
this condition, the parameters are set to the same values as those of the previous experiment except $\mu_{max}=3\times 10^{-3}$
for MASS and $\mu_{max}=5\times 10^{-3}$ for TAASS due to optimization. As can be seen from the figure, SINRs of all the
algorithms reduce at the same time. It is clear that the performance degradation of the proposed ones is much less significant
than those of the other methods. In addition, MASS and TAASS methods can quickly track the change and recover to a steady-state.
This figure illustrates that the proposed algorithms have faster convergence than the reported methods even though they are less
complex. The experiment shows that the proposed techniques exhibit better performance after an abrupt change, in a
non-stationary environment where the number of users/interferers suddenly changes in the system.

\section{\bfseries Conclusions}

In this paper, two novel adaptive step size mechanisms employing SG algorithms have been presented to enhance the performance,
improve the convergence property and reduce the computational load of the previously proposed adaptive methods for blind
adaptive beamforming. We considered different scenarios to compare the proposed MASS and TAASS algorithms with several existing
algorithms. Simulation experiments were conducted to investigate the output SINR. The performances of our new methods are shown
to be superior to those of others, both in terms of convergence rate and performance under sudden change in the signal
environment even though they are less complex. A complete convergence analysis of the proposed algorithms is under preparation.

\end{document}